\documentstyle[jkas]{article}

\beginpage{135}
\endpage{142}
\year{2011}\volume{44}\month{August}

\runningauthor {J. KIM ET AL.}
\runningtitle{NIR POLARIMETRY AROUND 30 DORADUS}

\month{August} \year{2011} \volume{44} \issueno{4}
\beginpage{135}\endpage{142}
\date{Received January 07, 2011; Revised August 01, 2011; Accepted August 02, 2011}

\begin{document}

  \title{
    NEAR-IR POLARIMETRY AROUND 30 DORADUS:\\
    I. SEPARATION OF THE GALACTIC SOURCES}

  \author{
  Jaeyeong Kim$^1$,
  Soojong Pak$^{1,\dag}$,
  Minho Choi$^2$,
  Wonseok Kang$^1$
  Ryo Kandori$^3$,
  Motohide Tamura$^3$,
  Tetsuya Nagata$^4$,
  Jungmi Kwon$^3$,
  Daisuke Kato$^5$, and
  Daniel T. Jaffe$^{1,6}$}

  \address{$^1$ School of Space Research, Kyung Hee University,
    1 Seocheon-dong, Giheung-gu, Yongin, Gyeonggi-do 446-701, Korea\\
    {\it E-mail : jaeyeong@khu.ac.kr, soojong@khu.ac.kr and wskang@khu.ac.kr}}
  \address{$^2$ Korea Astronomy and Space Science Institute,
    Daedeokdaero 776, Yuseong, Daejeon 305-348, Korea\\
    {\it E-mail : minho@kasi.re.kr}}
  \address{$^3$ National Astronomical Observatory of Japan, Mitaka,
    Tokyo 181-8588, Japan\\
    {\it E-mail : r.kandori@nao.ac.jp, motohide.tamura@nao.ac.jp and jmkwon@khu.ac.kr}}
  \address{$^4$ Department of Astronomy, Kyoto University,
    Kyoto 606-8502, Japan\\
    {\it E-mail : nagata@kusastro.kyoto-u.ac.jp}}
  \address{$^5$ Institute of Space and Astronomical Science, Japan Aerospace Exploration Agency,
    3-1-1 Yoshinodai, Chuo-ku, Sagamihara, Kanagawa, 252-5210, Japan\\
    {\it E-mail : kato@ir.isas.jaxa.jp}}
  \address{$^6$ Department of Astronomy, University of Texas at Austin,
    1 University Station, Austin, TX 78712-0259, USA\\
    {\it E-mail : dtj@astro.as.utexas.edu}}
  \address{\normalsize{\it (Received January 07, 2011; Revised August 01, 2011; Accepted August 02, 2011)}}

  \offprints{S. Pak}

\abstract{
A $20\arcmin \times 20\arcmin$ region around 30 Doradus in the Large Magellanic Cloud (LMC) is observed and analyzed in the near-infrared. We obtain polarimetry data in the $J$, $H$, and $K_s$ bands using the SIRIUS polarimeter SIRPOL at the Infrared Survey Facility 1.4 m telescope. We measure the Stokes parameters of 2562 point-like sources to derive the degree of polarization and the polarization position angles.
We discuss the statistics of the groups classified by color-magnitude diagram and proper motions of the sources, in order to separate the Galactic foreground sources from those present in the LMC. We notice that groups classified by the proper motion data show a tendency towards different polarimetric properties.}

\maketitle

\section{INTRODUCTION}
The role of magnetic fields is relevant in star formation processes and in galaxy evolution science (Kepley 2009; Schmidt 1970, 1976; Mathewson and Ford 1970; Wielebinski 1995). Measuring the orientation of dust grains is a way to investigate the physical properties of the magnetic field, since the magnetic field makes dust grains to align, due to rapid precession enforced from the substantial magnetic moment (Davis \& Greenstein 1951; Dolginov \& Mytrophanov 1976). In star-forming molecular clouds, dust grains tend to be aligned with their long axes perpendicular to the magnetic field (Lazarian 2006); dichroic polarization in the optical and near-IR wavelengths can measure the direction of those fields (Davis \& Greenstein 1951). In addition, the statistical pattern of the polarization position angles allows one to quantify the magnetic field strength (Ostriker et al. 2001; Kwon et al. 2010).

The Large Magellanic Cloud (LMC) is one of the nearest external galaxies. The interstellar medium of the LMC shows quite different physical and chemical properties with respect to that of our Galaxy, e.g., lower metallicity and different dust properties (Pak et al. 1998; Nakajima et al. 2007; Kim et al. 2010). The LMC has revealed large-scale magnetic field structures (Gaensler et al. 2005) from optical polarization observations (Schmidt 1970, 1976; Mathewson and Ford 1970; Wayte 1990). These features indicate that, in the west side of the LMC, the magnetic field is pointing towards the Small Magellanic Cloud. On the other hand, the magnetic fields around 30 Doradus appear to be complex structures. In fact, 30 Doradus is a unique, giant star formation complex in the LMC and contains massive clusters (Walborn \& Blades 1997; Brandner et al. 2001; Maercker \& Burton 2005; Scowen et al. 2009). Multi-frequency radio continuum surveys (Haynes et al. 1991) show the presence of filamentary magnetic fields in the LMC.

For this study, we observed near-infrared (NIR) sources in star forming regions ($\sim$39$\arcmin\times$69$\arcmin$) in the LMC. The number of detected sources, however, is quite large. In this paper, we only present preliminary results of selected regions around 30 Doradus ($\sim$20$\arcmin\times$20$\arcmin$). Section 2 describes the observations and the data reduction processes, including photometry and the derivation of the Stokes parameters. Section 3 includes the comparison of our data with previous observations by Nakajima et al (2007). In Section 4, we discuss the methods to extract Galactic foreground sources using the color-magnitude diagram and proper motions.

\section{OBSERVATIONS AND DATA REDUCTION}

The observations were carried out at the Infrared Survey Facility (IRSF) 1.4 m telescope in the South African Astronomical Obsevatory on December 25 and 30, 2008. Using the infrared camera SIRIUS (Nagayama et al. 2003) and the polarimeter SIRPOL (Kandori et al. 2006), we observed regions around 30 Doradus in \textit{J} (1.25 $\mu$m), \textit{H} (1.63 $\mu$m), and \textit{$K_s$} (2.14 $\mu$m) bands. This system enables us to obtain a wide-field (7$\farcm$7 $\times$ 7$\farcm$7) polarimetric image with a scale of 0$\farcs$45 pixel$^{-1}$.

We performed four 20-second exposures at wave-plate angles of 0$\arcdeg$, 45$\arcdeg$, 22$\arcdeg$.5, 67$\arcdeg$.5, and at 10 dithered positions for each field position. The total integration time was 200 seconds per wave-plate angle. The typical seeing size was $\sim$1$\farcs$3 in the $J$ band, but we experienced unstable seeing conditions during a night. Among the observed 5$\times$9 fields ($\sim$39$\arcmin\times$69$\arcmin$) we analyzed the central 3$\times$3 fields ($\sim$20$\arcmin\times$20$\arcmin$) and illustrate here our initial investigations (see Fig. 1). Table 1 lists the observation log of the data presented in this paper.

\begin{deluxetable}{ccccc}
\tablewidth{0pt}
\tablecaption{Observation Log \label{tbl-1}}
\tablehead{\colhead{Date(LT)} & \colhead{Field name} & \colhead{${{\alpha_{\circ\rm J2000.0}}}$} & \colhead{${\delta_{\circ\rm J2000.0}}$}
 & \colhead{Seeing in J band [$\arcsec$]}}\startdata

 2008 Dec 25 & n4 & 05 37 19.86 & -69 06 03.3 & 1.4\\
  & n5 & 05 38 34.48 & -69 06 02.2 & 1.4\\
  & n6 & 05 39 48.78 & -69 06 02.6 & 1.3\\
  & n7 & 05 39 48.17 & -69 12 38.0 & 1.3\\
  & n8 & 05 38 32.91 & -69 12 34.0 & 1.2\\
  & n9 & 05 37 17.79 & -69 12 32.4 & 1.3\\
  \hline\\
 2008 Dec 30 & n1 & 05 39 47.92 & -68 59 09.9 & 1.2\\
  & n2 & 05 38 33.53 & -68 59 07.7 & 1.3\\
  & n3 & 05 37 19.27 & -68 59 03.9 & 1.3\\
 \enddata
\end{deluxetable}
We used the IRAF\footnote{IRAF is distributed by the US National Optical Astronomy Observatories, which are operated by the Association of Universities for Research in Astronomy, Inc., under cooperative agreement with the National Science Foundation.} (Image Reduction \& Analysis Facility) software package to reduce the data, e.g, dark-field subtraction, flat-field correction and median sky subtraction. In addition, the data for each wave-plate position ($I_0$, $I_{45}$, $I_{22.5}$ and $I_{67.5}$) were processed as described in Kandori et al. (2006). For source detection and photometry on the Stokes $I$ images, we used the IRAF daophot package (Stetson 1987). We set the aperture radius to 8 pixels and the sky annulus to 13 pixels, and detected point sources having peak intensity greater than 10$\sigma$ above the local sky background. The pixel coordinates of the detected sources were matched to the celestial coordinates of their counterparts in the 2MASS\footnote{The Two Micron All Sky Survey(2MASS) is a joint project of the University of Massachusetts and the Infrared Processing and Analysis Center/California Institute of Technology.} Point Source Catalogue. The magnitude and color of our photometry were transformed into the 2MASS system using:

\begin{equation}
 m_{2MASS} = m_{IRSF} + \alpha_{1} \times C_{IRSF} + \beta_{1},\\\end{equation}
\begin{equation}
 C_{2MASS} = \alpha_{2} \times C_{IRSF} + \beta_{2},\\\end{equation}
where $m_{IRSF}$ is the instrumental magnitude from the IRSF images and $m_{2MASS}$ is the magnitude from the 2MASS All Sky Point Source Catalogue.

We calculated the Stokes parameters $I$, $Q$, and $U$, the polarization degree $P$, and the polarization position angle $\theta$, as follows:\\

\begin{equation}
I = (I_{0} + I_{22.5} + I_{45} + I_{67.5})/2,\\\end{equation}
\begin{equation}
Q = I_{0} - I_{45},\\\end{equation}
\begin{equation}
U = I_{22.5} - I_{67.5},\\\end{equation}
\begin{equation}
P_\circ = \frac{\sqrt{Q^{2}+U^{2}}}{I},\\\end{equation}
\begin{equation}
\theta = \frac{1}{2}\arctan{\frac{U}{Q}}.\\\end{equation}
Since the initial polarization degree $P_\circ$ is a positive quantity, the derived $P_\circ$ values tend to be overestimated, especially for low signal-to-noise ratio (S/N) sources. To correct for the bias, we use the following equation:
\begin{equation}
P = \sqrt{P_\circ^2-\delta P_\circ^2} ,\\\end{equation}
where $\delta P_\circ$ is the error in $P_\circ$ (Wardle \& Kronberg 1974). The polarization efficiencies of SIRPOL (95.5, 96.3, and 98.5\% at $J$, $H$, and $K_s$ bands, respectively) were also used for correcting the polarization degree (Kandori et al. 2006).

\begin{deluxetable}{ccccc}
\tablewidth{0pt}
\tablecaption{Statistical Properties of the Polarization angle}
\tablehead{\colhead{} & \colhead{$J^a$} & \colhead{$H^a$} & \colhead{${K_s}^a$} & \colhead{All bands$^a$}}\startdata

All sources & 79 $\pm$ 47 & 82 $\pm$ 48 & 85 $\pm$ 46 & 74 $\pm$ 24\\\\
\hline\\
$Group B$ & 75 $\pm$ 49 & 83 $\pm$ 53 & 75 $\pm$ 43 & 76 $\pm$ 11\\
$Group EFG$ & 81 $\pm$ 47 & 85 $\pm$ 47 & 87 $\pm$ 49 & 73 $\pm$ 21\\
\hline\\
$\mu$ $>$ 2$\sigma_\mu$ & 90 $\pm$ 45 & 102 $\pm$ 51 & 97 $\pm$ 44 & 96 $\pm$ 19\\
$\mu$ $<$ 2$\sigma_\mu$ & 69 $\pm$ 46 & 82 $\pm$ 48 & 83 $\pm$ 46 & 65 $\pm$ 24\\
\enddata
\tablecomments{$^a$ Polarization angle in degree units. The central angle and the sigma values are from obtained Gaussian fits to the distribution functions.}
\end{deluxetable}

\section{RESULTS}
From the data reduction process, we first derive polarization degrees and angles of 2562 sources, and extract only those with $P$ $\ge$ 3$\sigma_P$. ($\sigma_P$ indicates the error of a polarization value.) The numbers of the selected stars are 736, 784, and 427 in $J$, $H$, and $K_s$ bands, respectively. Fig. 2 shows a polarization vector map in the reduced 9 fields. The statistical distribution of the polarization degrees and angles is shown in Fig. 3.

To confirm our results, we compare our data with previous observations. Nakajima et al. (2007), hereafter N07, had observed the central 7$\farcm$7 $\times$ 7$\farcm$7 (one field) regions of 30 Doradus using the same SIRIUS/SIRPOL system. Our photometry in the 30 Doradus is limited by the nebula features, while N07 detected many point sources with long exposures, i.e., a total of 1480 seconds per wave-plate angle.

The numbers of the matched sources in both data are 33, 36, and 7, at $J$, $H$, and $K_s$ bands, respectively. With these sources, we compared the polarization degree and angle. Figure 4 shows that our values of the polarization angles and those from N07 are somehow consistent. However, the polarization degree values in our works are 50\% larger than those in N07. We suspect that this discrepancy is caused by the under-estimated $\delta P_\circ$ values in our low S/N data. More observations and analysis is required to confirm the calibration of the polarization degree values.

\section{DISCUSSION}
Source classification is important to separate Galactic foreground sources from those in the LMC, because they can contaminate polarimetry by their intrinsic polarization. In this section, we applied two methods to test the separation reliabilities.

\subsection{Color-Magnitude Diagram}
The typical classification method is to use color-magnitude diagrams (hereafter $CM$ diagrams). Nikolaev \& Weinberg (2000), hereafter NW00, classified the LMC stars using 2MASS data as follows:\\
$\bullet\ Group$ $A$: blue supergiants and O dwarfs in the LMC;\\
$\bullet\ Group$ $B$: Galactic disk dwarfs with spectral classes of F$-$K and Galactic giants with spectral classes of F$-$G;\\
$\bullet\ Group$ $D$: Galactic disk dwarfs with spectral classes of G$-$M, red-giant branch (RGB), and early asymptotic giant branch (E-AGB) stars in the LMC;\\
$\bullet\ Group$ $EFG$: tip of the RGB, E-AGB, and O-rich AGB stars in the LMC ;\\
$\bullet\ Group$ $I$: super giants and the red clump in the LMC;\\
$\bullet\ Group$ $JK$: C-rich AGB and dusty AGB stars in the LMC.\\

The 2MASS sources, however, are limited in $m_{K_s}$ $<$ 14 mag. On the other
hand, our photometry data cover up to $m_{K_s}$ $<$ 16 mag and most detected
sources are populated under $Group$ $D$ in NW00. Kato et al. (2007) used the
criteria of NW00, but did not classify the sources which distributed
under $Group$ $D$. We followed the same criteria as Kato et al. (2007)
to separate sources. Figure 5 shows our five groups.

Galactic foreground stars are mostly distributed above the brighter region ($m_{K_s}$ $<$ 11) of $Group$ $A$ according to NW00. However, our data do not contain the detected sources in that region. The detected foreground sources are distributed in $Group$ $B$; the Galactic giants with spectral classes of F$-$G have a small population ($\le$ 5\%) within this group. Most RGB and AGB stars in the LMC are in $Group$ $EFG$. Note that $Group$ $A$ and $Group$ $I$ can be confused with the Galactic foreground sources ($Group$ $B$) because of their uncertain boundary lines. $Group$ $JK$ have dusty AGB stars which may have their own polarization by the shrouded dust shells (Parthasarathy \& Jain 1993). Therefore, $Group$ $A$, $Group$ $I$ and $Group$ $JK$ are excluded from the samples when comparing the foreground and the main LMC sources.

We classify $Group$ $B$ as foreground sources and $Group$ $EFG$ as the main LMC stars. We extrapolate the boundaries of the criteria regions up to $m_{K_s}$ $<$ 16 mag and plot histograms for the sources in $Group$ $B$ and $Group$ $EFG$. Figure 6 and 7 show the distributions of polarization degrees and angles for both $Group$ $B$ and $Group$ $EFG$. In Figure 6, the polarization degree distribution of $Group$ $EFG$ shows a large population of high polarization values ($P$ $>$ 10$\%$) which comes from the intrinsic polarization of the RGB and AGB stars. On the other hand, the polarization angle distributions show similar tendency. See the statistical distributions in Figure 7 and Table 2. Because the criteria of $Group$ $B$ and $Group$ $EFG$ are overlapped in 14 $<$ $m_{K_s}$ $<$ 16, we expect that many sources may not be classified properly.

\subsection{Proper Motion}
Vieira et al. (2010) investigated the proper motion of the Magellanic Clouds from the resulting catalog of 1.4 million objects in the Southern Proper Motion (SPM) program. They used the proper motion of 3822 and 964 sources near the clouds to derive the mean absolute proper motions of the LMC and of the SMC.

We used this catalog and examined proper motions to separate the foreground sources which would be contained in our observed regions in the LMC. A total of 805 sources were matched with SPM sources in the $JHK_s$ bands. Vieira et al. (2010) discarded areas of high stellar density, at $4^h 40^m 58.8^s$ $\le$ $\alpha$ $\le$ $5^h 40^m 12^s$ and $-68\arcdeg$ $\le$ $\delta$ $\le$ $-71\arcdeg$, to avoid the misidentified measurement. Since our observed regions are located on the edge of the discarded area, some of our data may not be reliable.

In order to reduce the uncertainty in the data, we used data which had proper motions with $\mu$ $>$ 2$\sigma_\mu$ to select Galactic foreground stars. Fig. 8 shows the distribution of the sources according to the S/N of the proper motion. Based on the existence of proper motion, we examined the difference between the LMC sources and the Galactic foreground ones.

Figure 9 shows that the histograms of the polarization degrees are similar between the two groups, while those of the $CM$ diagram classifications in Figure 6 are very different. We expect that the difference in Figure 6 is partly from the intrinsic differences of the spectral types. As evident from Figure 10, the polarization angle distributions are slightly different. The Galactic foreground group with proper motions shows almost equally distributed angles with a slight peak at 93$\arcdeg$, while that of the LMC group shows a peak at 75$\arcdeg$.

We also examined the possibility that the Galactic foreground stars may show some polarization properties different from the main LMC stars by their intrinsic polarization or the effect for the Galactic magnetic fields. Previous observations (Mathewson \& Ford 1970; Isserstedt \& Reinhardt 1976; Schmidt 1976) indicated that the Galactic magnetic fields at the line of sight of the LMC can influence the polarization properties of the foreground stars. According to those papers, the polarization properties of the Galactic sources in front of the LMC have a 0.3\% polarization degree and 30\arcdeg of polarization angle. However, these results may not be applied to ours because our detection limits are too low to measure the dichroic polarization of the Galaxy.

\section{SUMMARY}
We conducted imaging polarimetry of the 3$\times$3 fields ($\sim$20$\arcmin\times$20$\arcmin$) around 30 Doradus. Because of unstable seeing condition, our aperture radius for photometry and polarimetry has been set 2--3 times larger than the usual aperture radius. We used some previous data to confirm our results. Comparing with the sources of N07, we found no significant differences in our polarimetry data.

We applied several methods to remove the Galactic foreground sources which can contaminate the polarimetry result by their intrinsic polarization. Using the $CM$ diagram, we selected the group which contains the foreground sources. The main LMC sources show larger polarization degrees than the foreground sources.

We also selected the Galactic foreground sources using proper motion data. The foreground sources show high values of proper motion, contrary to the main LMC sources.

Comparing different proper motions, we can separate out the Galactic foreground sources from the main LMC sources. They showed different distributions for polarization degrees and angles with respect to the main LMC sources.

\acknowledgments
{J. Kim was supported by a WCU (World Class University) program through
the National Research Foundation of Korea funded by the Ministry of Education,
Science and Technology  (R31-10016). Part of this work was supported by the National Research Foundation of Korea funded by the Ministry of Education, Science and Technology No. 2009-0063616.}

\clearpage
\begin{figure*}[p]
\centering
\epsfxsize=6cm
\epsfbox{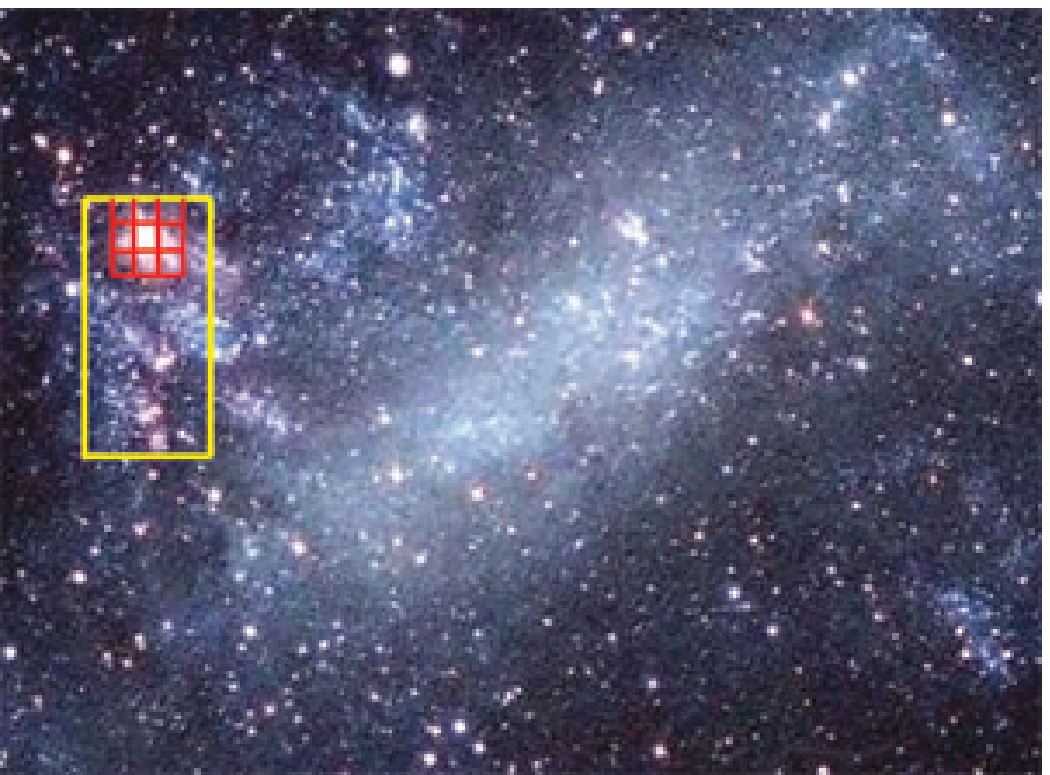}
\caption{Observed area of the LMC on an optical image offered by Mr. Kamiya. The yellow box shows the total 5$\times$9 survey field by SIRPOL on 2008 December 25 and 30 and the red box is the 3$\times$3 fields around 30 Doradus.}
\end{figure*}
\begin{figure*}[p]
\centering
\epsfxsize=16cm
\epsfbox{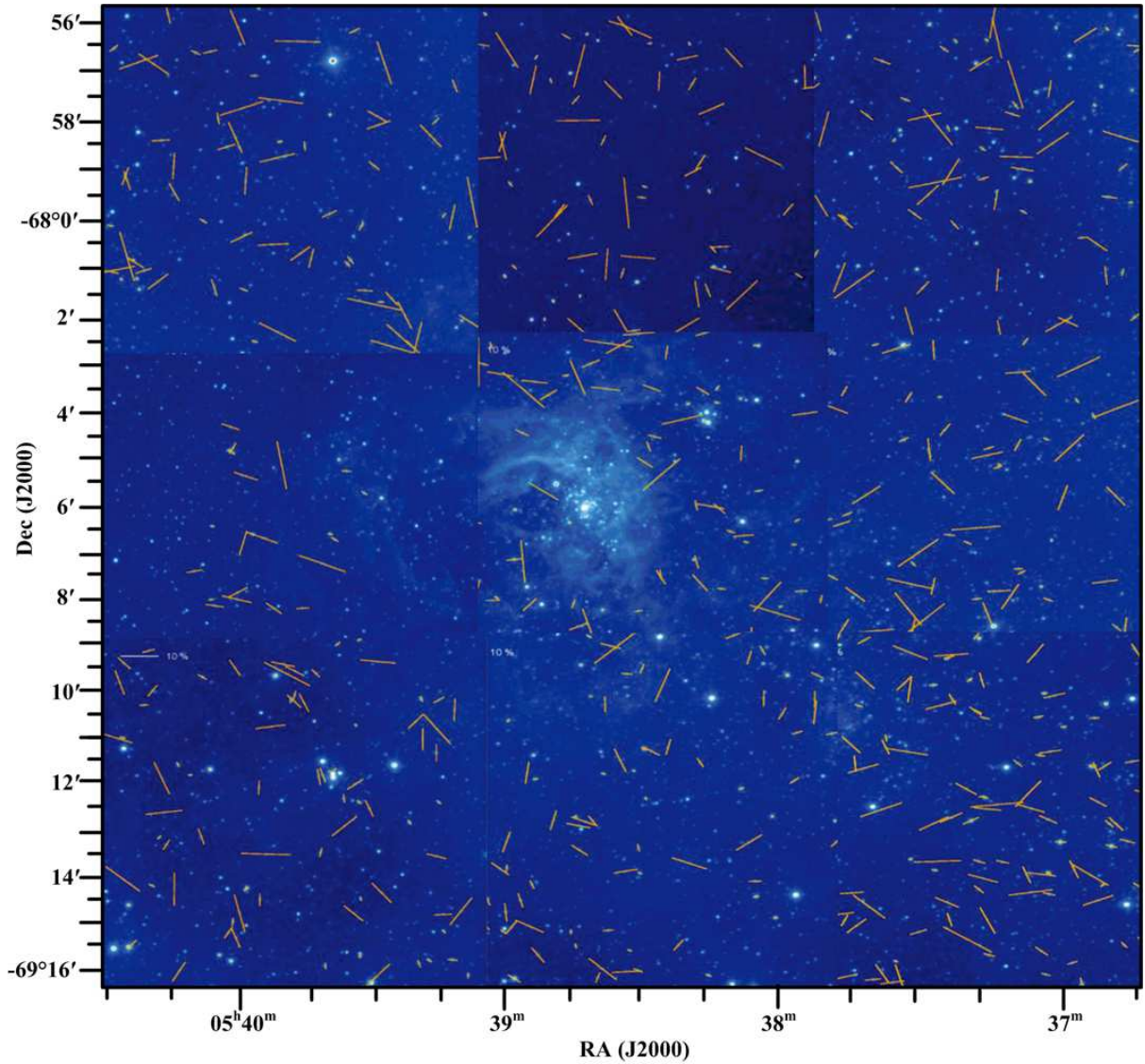}
\caption{$H$ band intensity image of the 3$\times$3 fields and polarization vectors (orange lines) of the polarized
 sources in the $H$ band. The length of the vectors indicates the polarization degree.}
\end{figure*}
\begin{figure*}[p]
\centering
\epsfxsize=12cm
\includegraphics[scale=0.4]{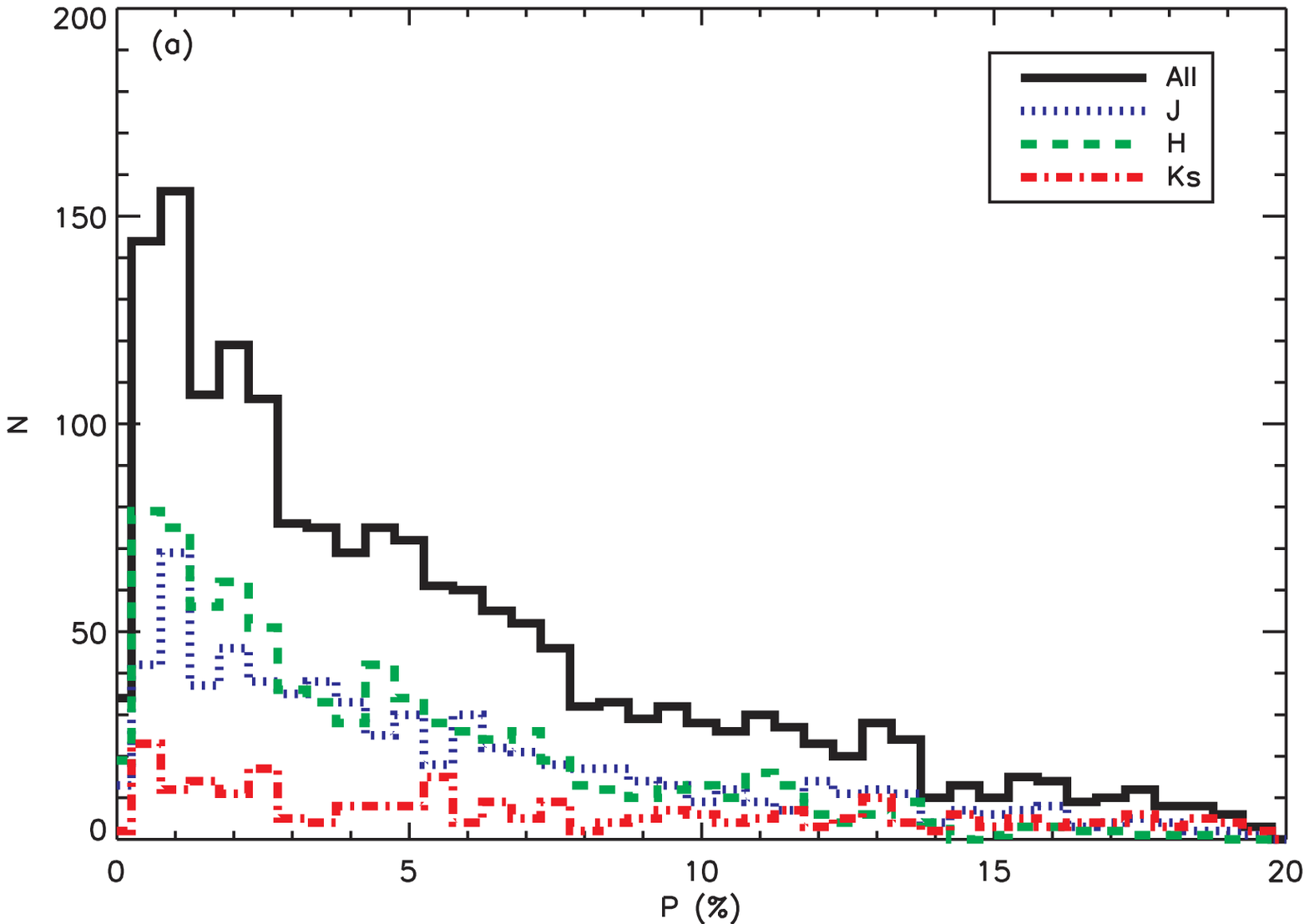}\\
\includegraphics[scale=0.4]{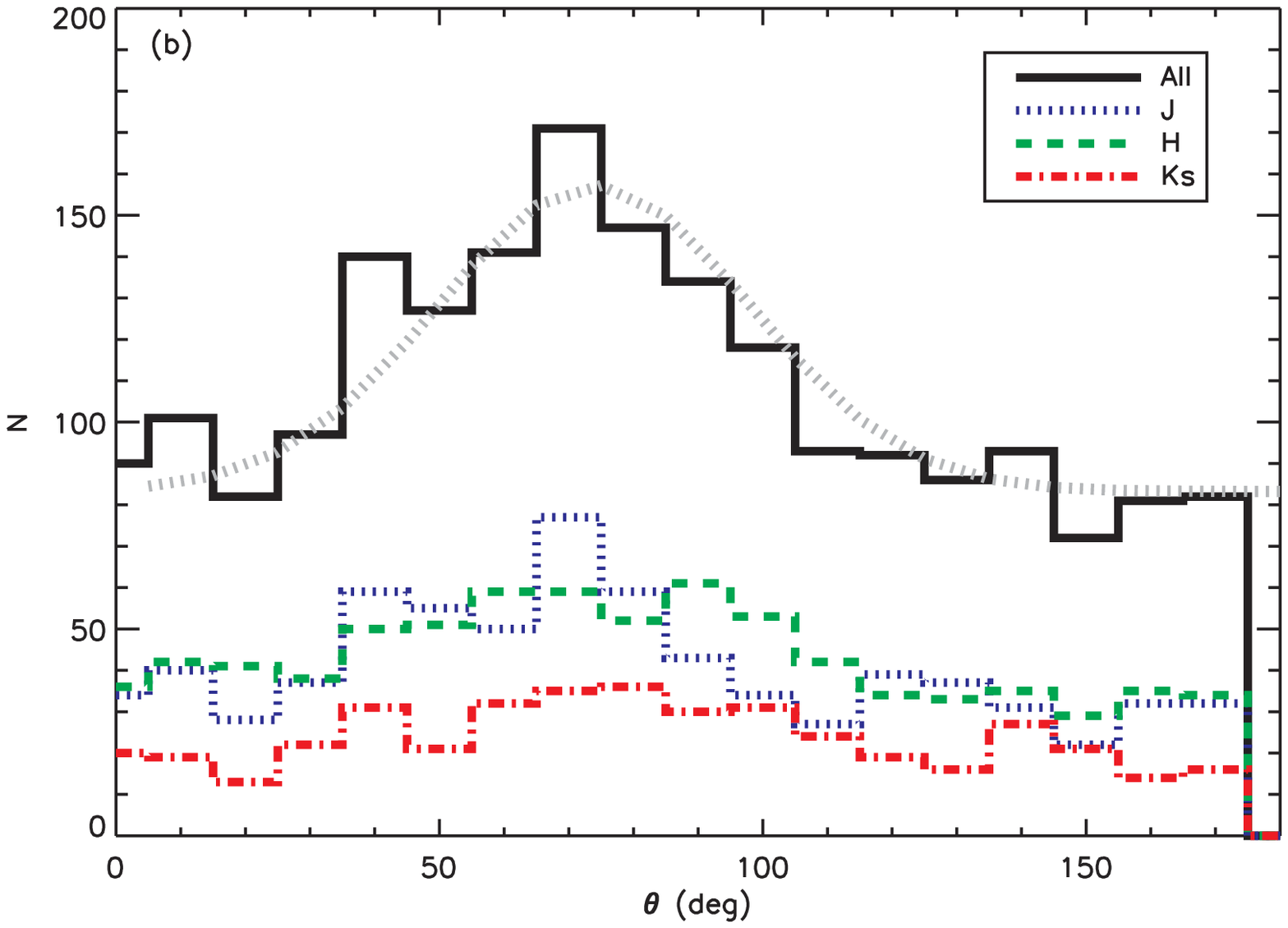}
\caption{(a) Histograms of polarization degree of the sources in nine fields with $P$ $\ge$ 3$\sigma_P$ for the $J$, $H$, and $K_s$ bands. (b) Histograms of polarization angle of the sources in nine fields with $P$ $\ge$ 3$\sigma_P$ for the $J$, $H$, and $K_s$ bands. The grey dotted line on the histogram of the all sources show the results of Gaussian fits. The central angle and the sigma of the Gaussian function are listed in Table 2.}
\end{figure*}
\begin{figure*}[p]
\centering
\epsfxsize=12cm
\includegraphics[scale=0.5]{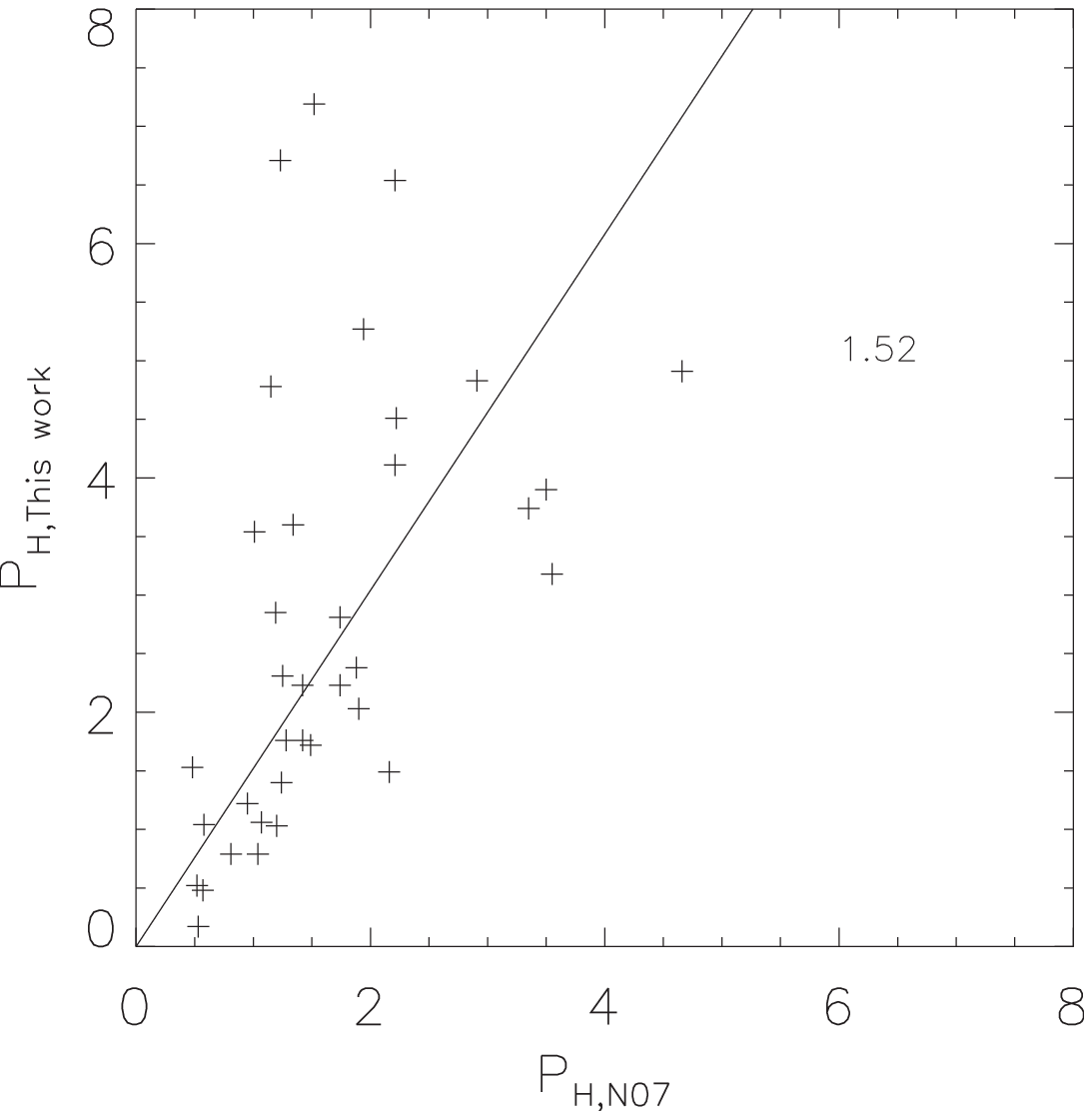}
\includegraphics[scale=0.5]{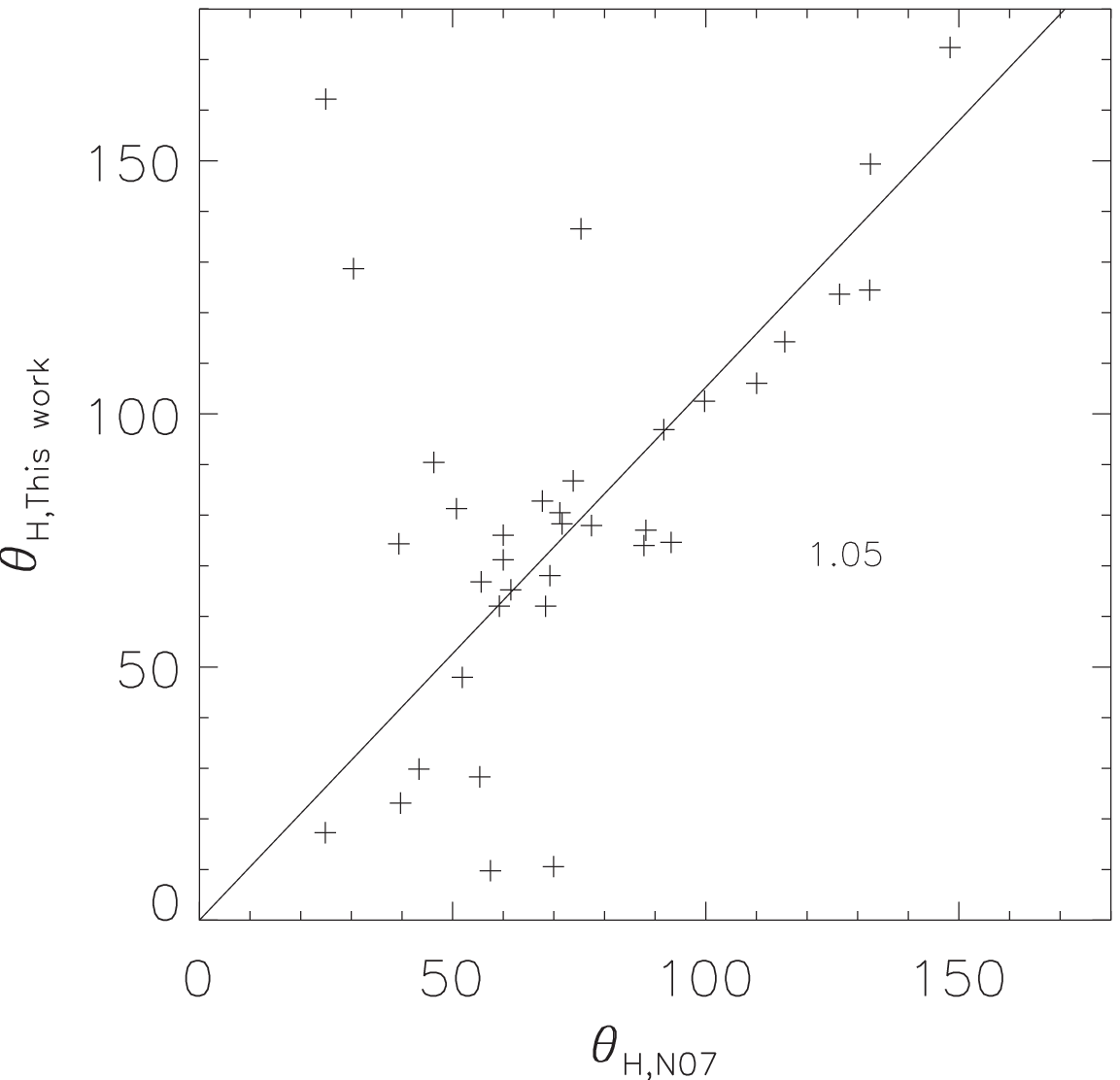}\\
\caption{Correlation graph of matched data in the $H$ band.
Polarization degree (left) and angle (right) are shown. The solid lines are linear fits with slopes of 1.52 (left) and 1.05 (right).}
\end{figure*}
\begin{figure*}[p]
\centering
\epsfxsize=12cm
\includegraphics[scale=0.5]{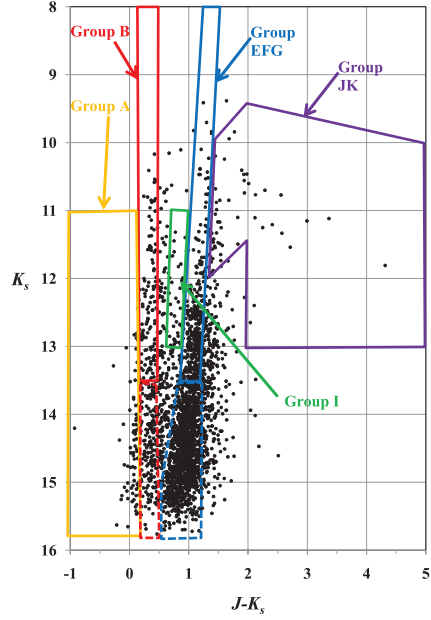}
\caption{Color-magnitude ($J-K_s$ vs. $K_s$) diagram of our sources. Each group is indicated with a different color box: $Group$ $A$ (orange box), $Group$ $EFG$ (blue box), $Group$ $B$ (red box), $Group$ $I$ (green box), and $Group$ $JK$ (violet box). Our photometry data can detect more faint sources ($m_{K_s}$ $<$ 16 mag) than the previous data of Nikolaev \& Weinberg (2000) and most of the sources are distributed in the range of 14 $<$ $m_{K_s}$ $<$ 16 mag. The red and blue dashed boxes are extrapolated regions from $Group$ $B$ and $Group$ $EFG$.}
\end{figure*}
\begin{figure*}[p]
\centering
\epsfxsize=12cm
\includegraphics[scale=0.4]{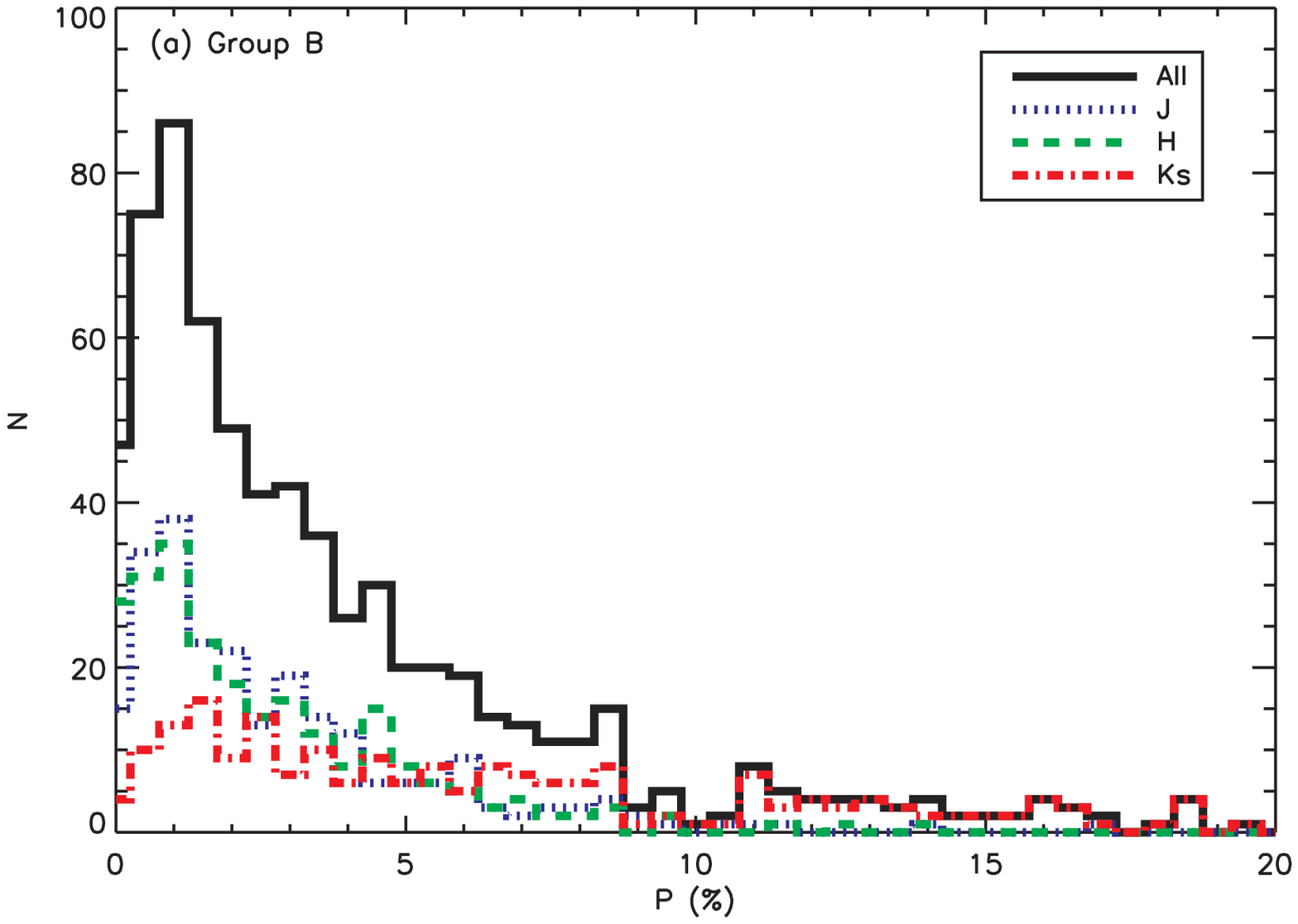}\\
\includegraphics[scale=0.4]{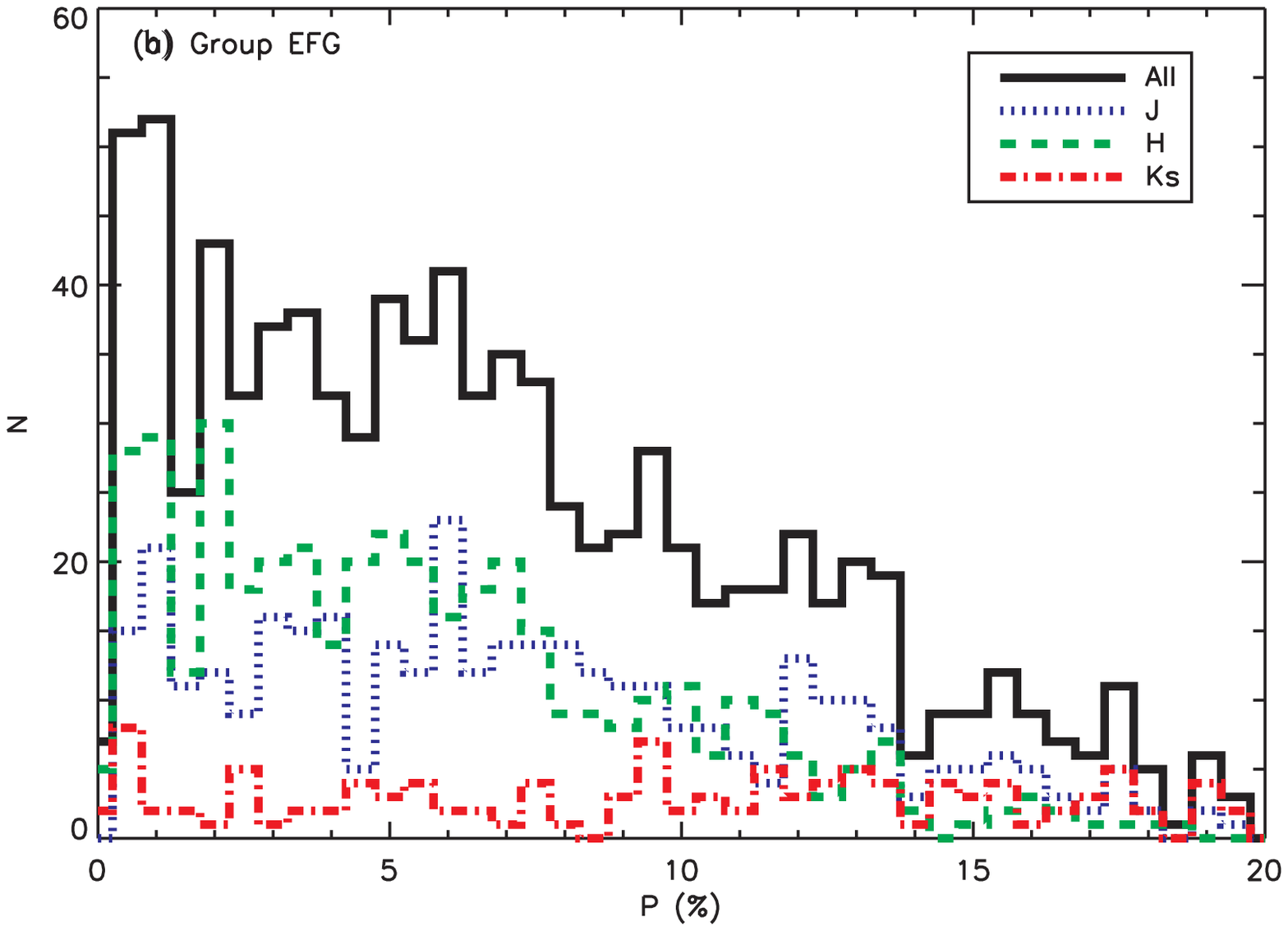}
\caption{(a) Histograms of the polarization degree of the sources which are classified to the $Group$ $B$ for the $J$, $H$, and $K_s$ bands. (b) Histograms of the polarization degree of the sources which are classified to the $Group$ $EFG$ and $P$ $\ge$ 3$\sigma_P$ for the $J$, $H$, and $K_s$ bands.}
\end{figure*}

\begin{figure*}[p]
\centering
\epsfxsize=12cm
\includegraphics[scale=0.4]{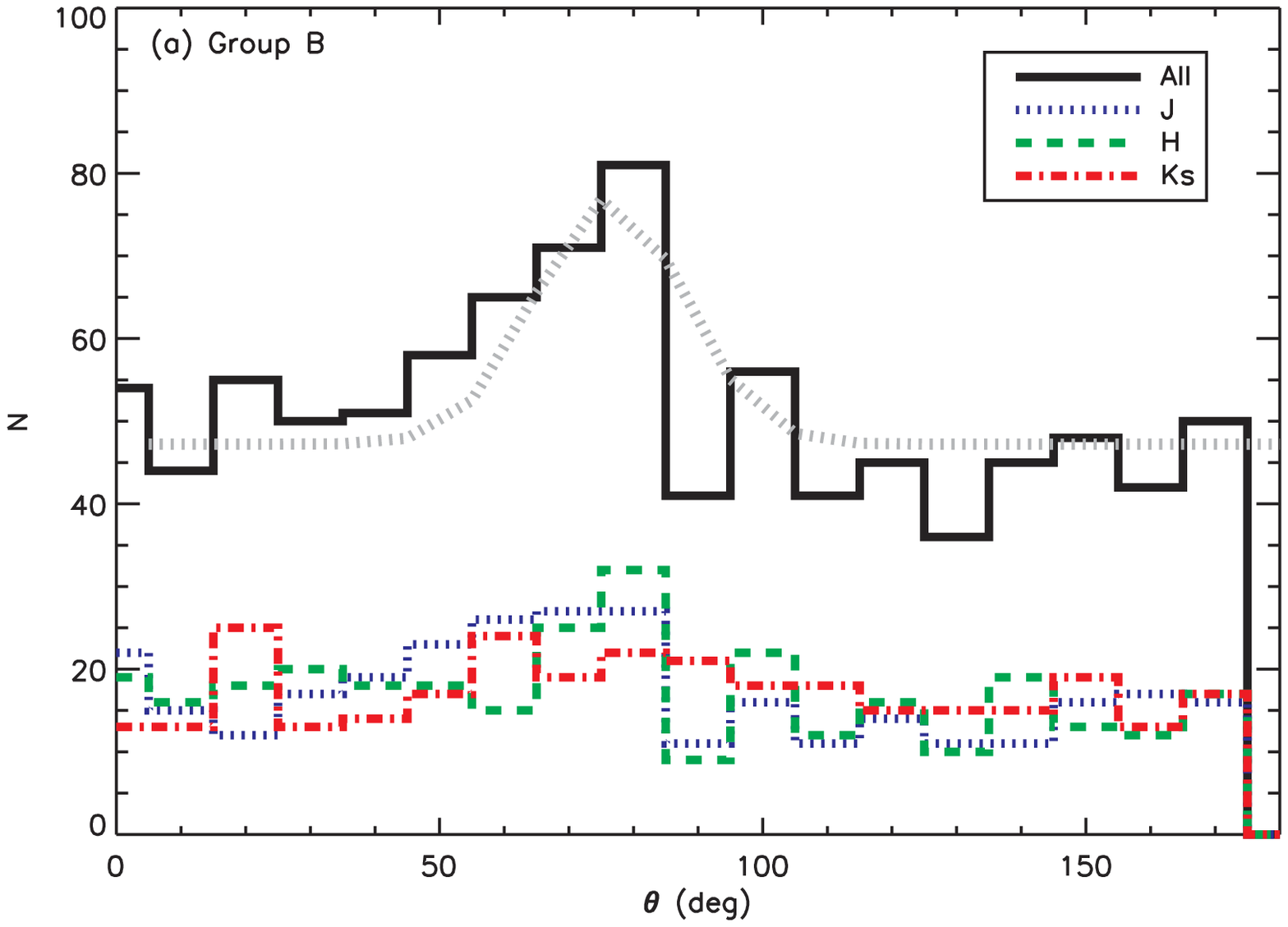}\\
\includegraphics[scale=0.4]{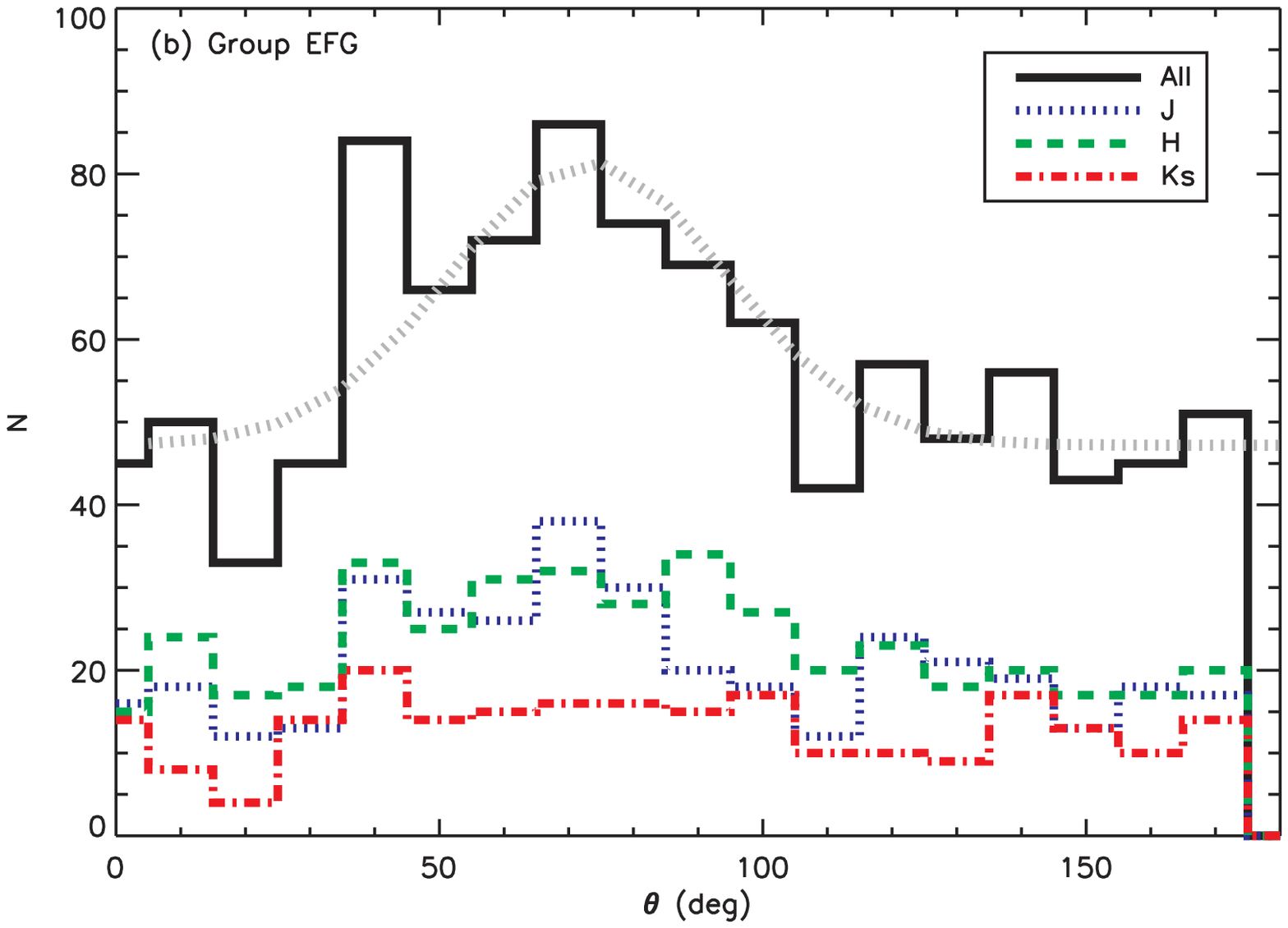}
\caption{(a) Histograms of the polarization angle of the sources which belong to the $Group$ $B$ for the $J$, $H$, and $K_s$ bands. (b) Histograms of the polarization angle of the sources which belong to the $Group$ $EFG$ and $P$ $\ge$ 3$\sigma_P$ for the $J$, $H$, and $K_s$ bands.}
\end{figure*}
\begin{figure*}[p]
\centering
\epsfxsize=12cm
\includegraphics[scale=0.4]{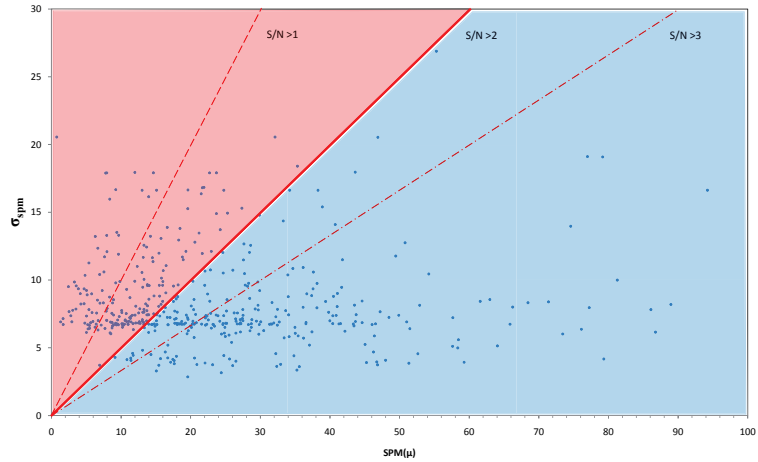}
\caption{Plots of the proper motion and of its error. The blue region (S/N $>$ 2) may contain most of the foreground sources, and the main LMC sources are mostly distributed in the red region (S/N $<$ 2). From the S/N ratio of the proper motion we try to separate sources.}
\end{figure*}
\begin{figure*}[p]
\centering
\epsfxsize=12cm
\includegraphics[scale=0.4]{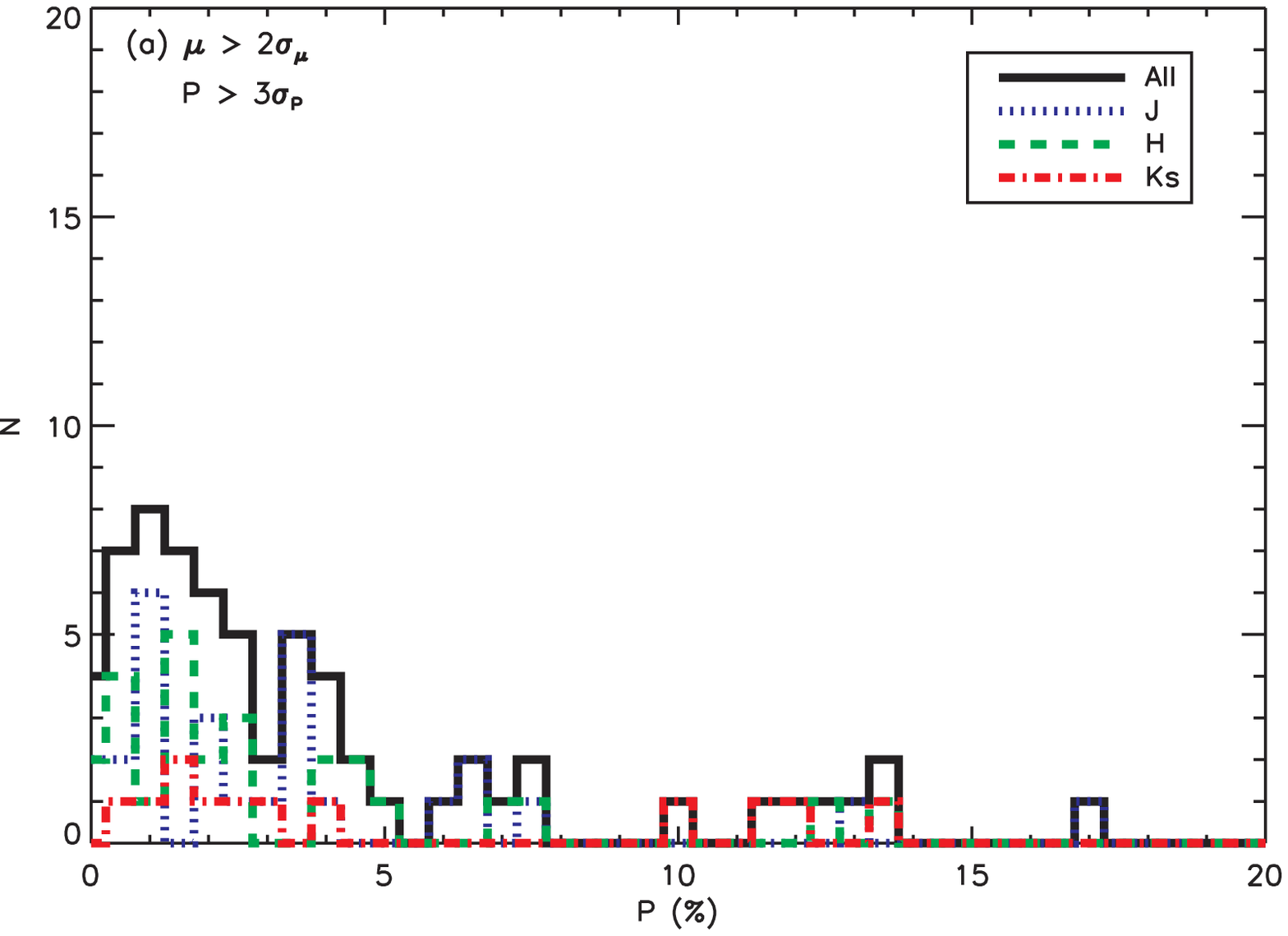}\\
\includegraphics[scale=0.4]{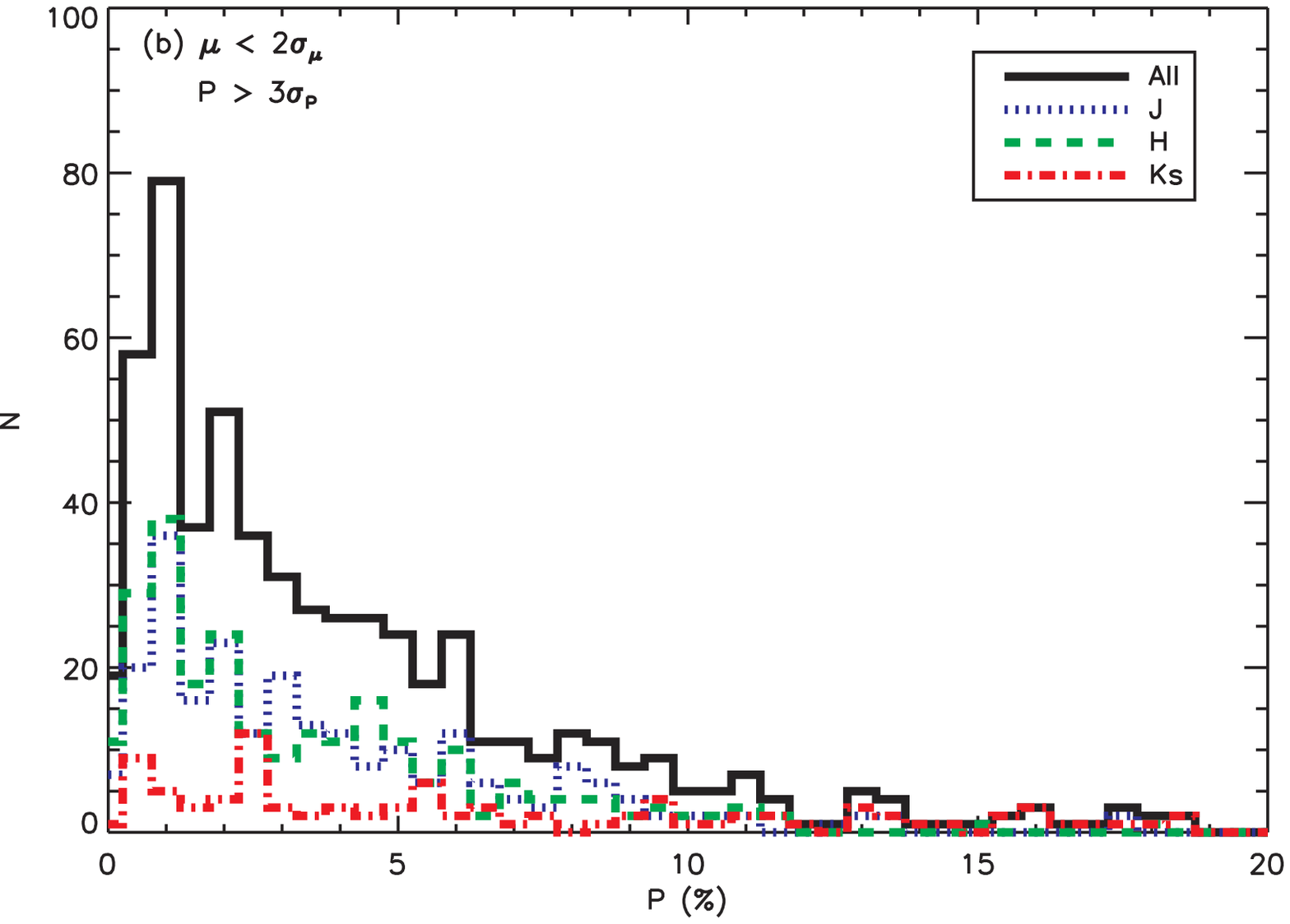}
\caption{(a) Histograms of the polarization degree of the sources which have proper motion $\mu$ $>$ 2$\sigma_\mu$ and $P$ $>$ 3$\sigma_P$ for the $J$, $H$, and $K_s$ bands. (b) Histograms of the polarization degree of the sources which have proper motion $\mu$ $<$ 2$\sigma_\mu$ and $P$ $>$ 3$\sigma_P$ for the $J$, $H$, and $K_s$ bands.}
\end{figure*}
\begin{figure*}[p]
\centering
\epsfxsize=12cm
\includegraphics[scale=0.4]{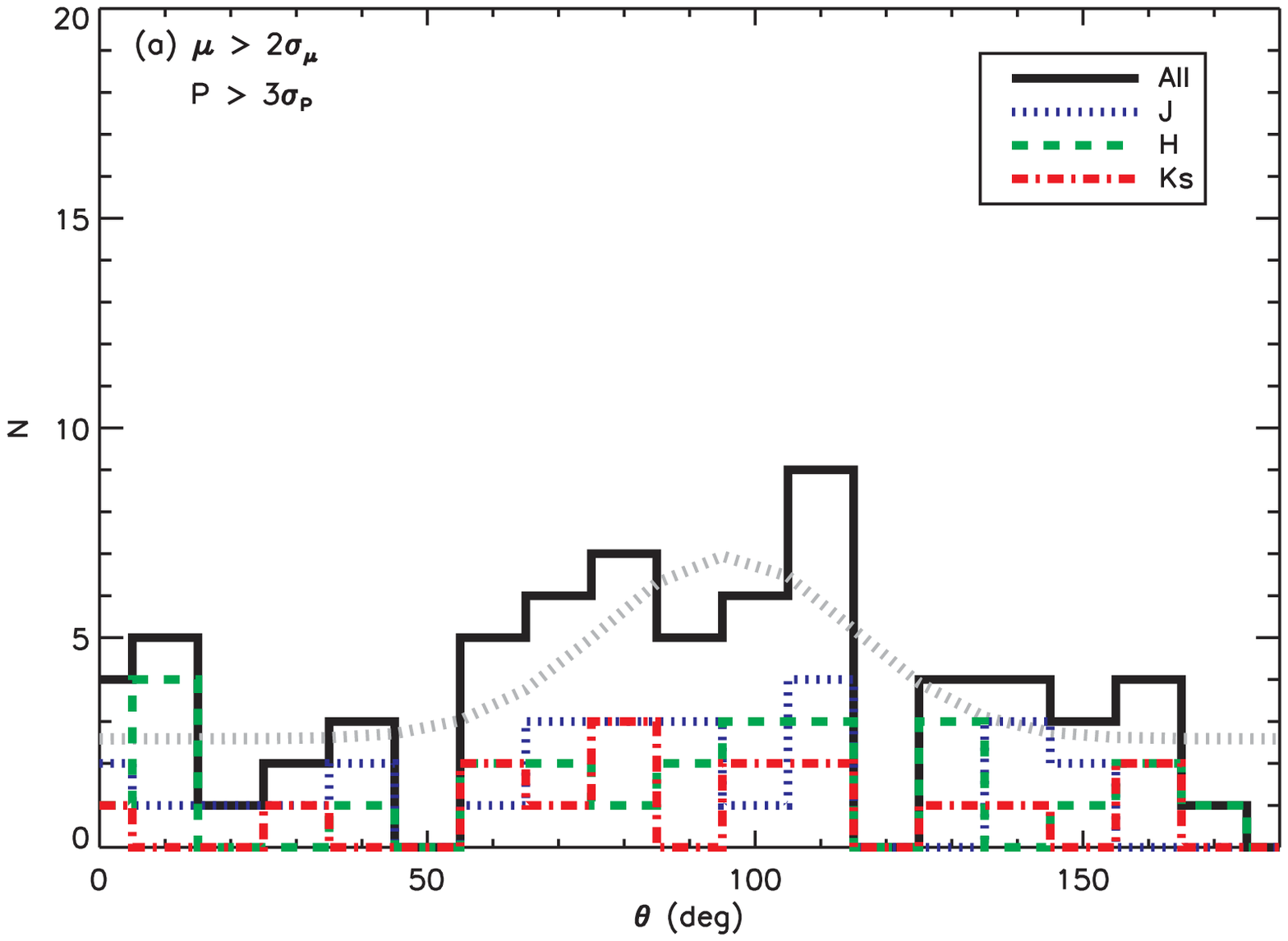}\\
\includegraphics[scale=0.4]{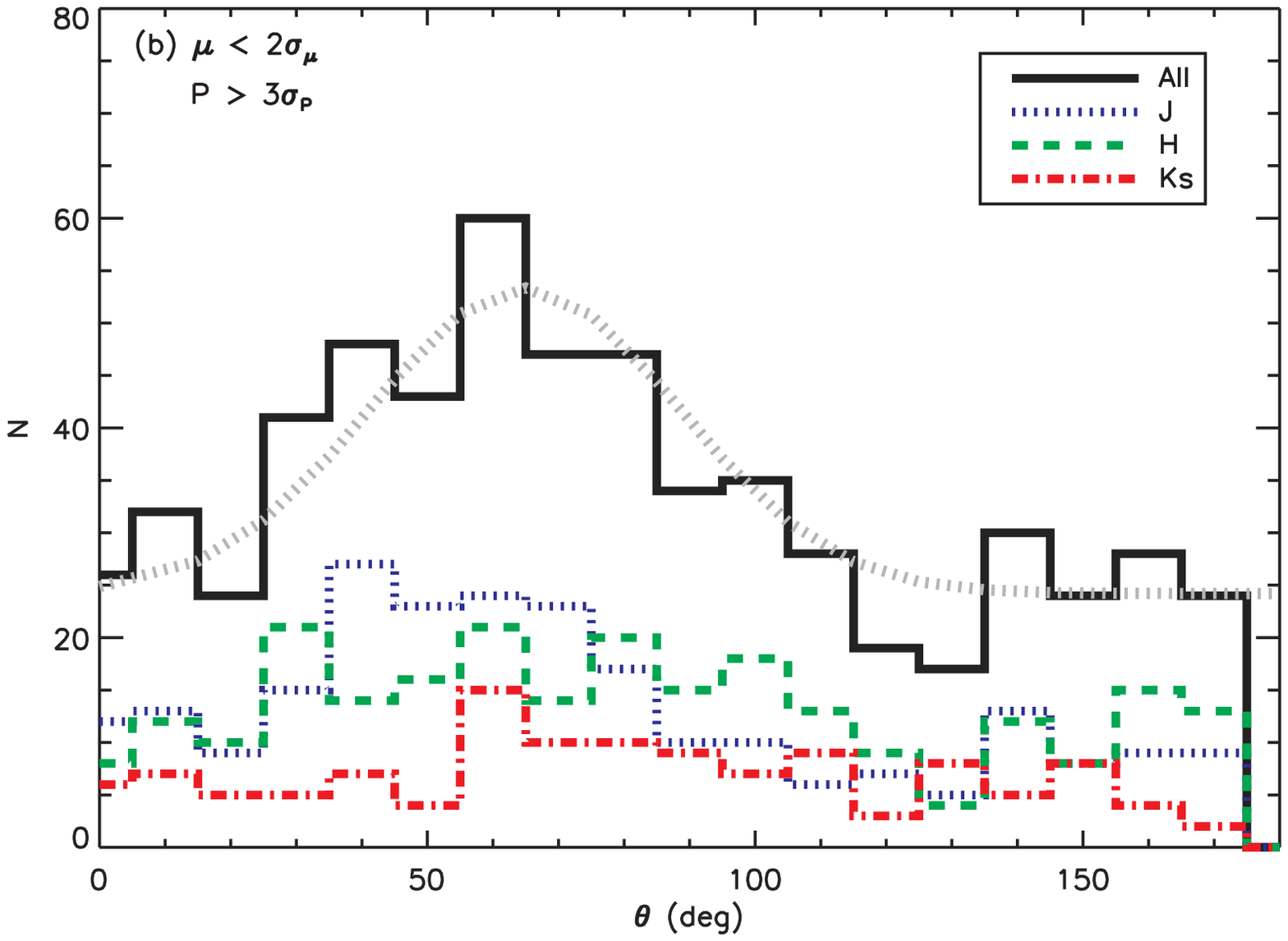}
\caption{(a) Histograms of the polarization angle of the sources which have proper motion $\mu$ $>$ 2$\sigma_\mu$ and $P$ $>$ 3$\sigma_P$ for the $J$, $H$, and $K_s$ bands. (b) Histograms of the polarization angle of the sources which have proper motion $\mu$ $<$ 2$\sigma_\mu$ and $P$ $>$ 3$\sigma_P$ for the $J$, $H$, and $K_s$ bands.}
\end{figure*}


\begin{thebibliography}{}
\bibitem[Brandner et al. (2001)]{brandner01} Brandner, W., Grebel, E. K., Barb\'{a}, R. H., Walborn, N. R., \& Moneti, A. 2001, Hubble Space Telescope NICMOS Detection of a Partially Embedded, Intermediate-Mass, Pre-Main-Sequence Population in The 30 Doradus Nebula, \aj, 122, 858
\bibitem[Davis \& Greenstein(1951)]{davis51} Davis, L. J., \& Greenstein, J. L. 1951, The Polarization of Starlight by Aligned Dust Grains, \apj, 114, 206
\bibitem[Dolginov \& Mytrophanov(1976)]{dolginov76} Dolginov, A. Z., \& Mytrophanov, I. G. 1976, Orientation of Cosmic Dust Grains, \apss, 43, 291
\bibitem[Gaensler et al.(2005)]{gaensler05} Gaensler, B. M., Haverkorn, M., Staveley-Smith, L., Dickey, J. M., McClure-Griffiths, N. M., Dickel, J. R., \& Wolleben, M. 2005, The Magnetic Field of the Large Magellanic Cloud Revealed through Faraday Rotation, Science, 307, 1610
\bibitem[Haynes et al.(1991)]{haynes91} Haynes, R. F., et al. 1991, A Radio Continuum Study of the Magellanic Clouds, \aap, 252, 475
\bibitem[Isserstedt \& Reinhardt(1976)]{isserstedt76} Isserstedt, F., \& Reinhardt, M. 1976, On The Structure of The Magnetic Field in The Large Magellanic Cloud, MNRAS, 176, 693
\bibitem[Kandori et al.(2006)]{kandori06} Kandori, R., et al. 2006, SIRPOL: a JHKs-Simultaneous Imaging Polarimeter for the IRSF 1.4-m Telescope, \procspie, 6269, 159
\bibitem[Kandori et al.(2007)]{kandori07} Kandori, R., et al. 2007, Near-Infrared Imaging Polarimetry of the Star-Forming Region NGC 2024, \pasj, 59, 487
\bibitem[Kato et al.(2007)]{kato07} Kato, D., et al. 2007, The IRSF Manellanic Clouds Point Source Catalog, \pasj, 59, 615
\bibitem[Kepley et al.(2009)]{kepley07} Kepley, A. A., M\"{u}hle S., Wilcots, E. M., Everett, J., Zweibel, E., Robishaw, T., \& Heiles, C. 2007, Magnetic Fields in Irregular Galaxies, IAUS, 259, 555
\bibitem[Kim et al.(2010)]{kim10} Kim, S., et al. 2010, Colod Dust Clumps in Dynamically Hot Gas, \aap, 518, L75
\bibitem[Kwon et al.(2010)]{kwon10} Kwon, J., Choi, M., Pak, S., Kandori, R., Tamura, M., Nagata, T., \& Sato, S. 2010, Magnetic Structure of The HH 1-2 Region: Near-Infrared Polarimetry of Point-Like Sources, \apj, 708, 758
\bibitem[Lazarian(2006)]{lazarian06} Lazarian, A. 2006, Grain Alignment, Polarization and Magnetic Fields, \aaps, 8049
\bibitem[Maercker \& Burton(2005)]{maercker05} Maercker, M., \& Burton, M. G. 2005, L-band (3.5$\mu$m) IR-excess in massive star formation I. 30 Doradus, \aap, 438, 663
\bibitem[Mathewson \& Ford(1970)]{methewson70} Mathewson, D. S., \& Ford, V. L. 1970, The Magnetic-Field Structure of the Magnellanic Clouds, \apj, 160, L43
\bibitem[Nakajima et al.(2007)]{nakajima07} Nakajima, Y., et al. 2007, First NIR Polarimetry of 30 Doradus, \pasj, 59, 519
\bibitem[Nagayama et al.(2003)]{nagayama03} Nagayama, T., et al. 2003, SIRIUS: a Near Infrared Simultaneous Three--Band Camera, \procspie, 4841, 459
\bibitem[Nikolaev \& Weinberg(2000)]{nikolaev00} Nikolaev, S., \& Weinberg, M. D. 2000, Stellar Populations in the Large Magellanic Cloud from 2MASS, \apj, 542, 804
\bibitem[Ostriker et al.(2001)]{ostriker01} Ostriker, E. C., Stone, J. M. \& Gammie, C. F. 2001, Density, Velocity, and Magnetic Field Structure in Turbulent Molecular Cloud Models, \apj, 546, 980
\bibitem[Pak et al.(1998)]{pak98} Pak, S., Jaffe, D. T., van Dishoeck, Ewine F., Johansson, L. E. B., \& Booth, R. S. 1998, Molecular Cloud Structure in the Magellanic Clouds: Effect of Metallicity, \apj, 498, 735
\bibitem[Parthasarathy \& Jain(1993)]{parthasarathy93} Parthasarathy, M., \& Jain, S. K. 1993, UNVRI Polarization Measurements of POST AGB Stars, IAUS, 155, 353
\bibitem[Schmidt(1970)]{schmidt70} Schmidt, Th. 1970, Polarization Measurements and Magnetic Field Structure within the Magellanic Clouds, \aap, 6, 294
\bibitem[Schmidt(1976)]{shmidt76} Schmidt, Th. 1976, Starlight Polarization in the Magellanic Cloud Regions, \aap, 24, 357
\bibitem[Scowen et al.(2009)]{scowen09} Scowen, P., et al. 2009, The Magellanic Clouds Survey: a Bridge to Nearby Galaxies, Astro, 266
\bibitem[Stetson(1987)]{stetson87} Stetson, P. B. 1987, DAOPHOT: A Computer Program for Crowded--Field Stellar Photometry, \pasp, 99, 191
\bibitem[Tinbergen(1996)]{tinbergen96} Tinbergen, J. 1996, Dichroism, in Astronomical Polarimetry (Boston: Cambridge Univ. press), 42
\bibitem[Vieira et al.(2010)]{vieira10} Vieira, K., et al. 2010, Proper Motion Study of the Magellanic Clouds using SPM material, \aj, 140, 1934
\bibitem[Wardle \& Kronberg(1974)]{wardle74} Wardle, J. F. C., \& Kronberg, P. P. 1974, The Linear Polarization of Quasi--Stellar Radio Sources at 3.71 and 11.1 centimeters, \apj, 194, 249
\bibitem[Walborn \& Blades(1997)]{walborn97} Walborn, N. R., \& Blades J. C. 1997, Spectral Classification of The 30 Doradus Stellar Populations, \apjs, 112, 457
\bibitem[Wayte(1990)]{wayte90} Wayte, S. R. 1990, Structure of the Interstellar Medium in the Magellanic Clouds, \apj, 355, 473
\bibitem[Wielebinski(1995)]{wielebinski95} Wielebinski, R. 1995, Galactic and Extragalactic Magnetic Fields, Reviews in Modern Astronomy, Vol. 8, Reviews in Modern Astronomy, ed. G. Klare, 185-200

\end{thebibliography}
\end{document}